\documentstyle[12pt,epsfig]{article}

\newlength{\dinwidth}
\newlength{\dinmargin}
\begin{document}
\def\bold#1{\setbox0=\hbox{$#1$}%
     \kern-.025em\copy0\kern-\wd0
     \kern.05em\%\baselineskip=18ptemptcopy0\kern-\wd0
     \kern-.025em\raise.0433em\box0 }
\def\slash#1{\setbox0=\hbox{$#1$}#1\hskip-\wd0\dimen0=5pt\advance
         to\wd0{\hss\sl/\/\hss}}
\newcommand{\be}{\begin{equation}}
\newcommand{\ee}{\end{equation}}
\newcommand{\bea}{\begin{eqnarray}}
\newcommand{\eea}{\end{eqnarray}}
\newcommand{\nn}{\nonumber}
\newcommand{\dd}{\displaystyle}
\newcommand{\bra}[1]{\left\langle #1 \right|}
\newcommand{\ket}[1]{\left| #1 \right\rangle}
\newcommand{\spur}[1]{\not\! #1 \,}
\def \de {\partial}
\def \l {\lambda}
\def \L {\Lambda}
\def \T {\Theta}
\def \X {\Xi}
\def \p {\pi}
\def \P {\Pi}
\def \D {\Delta}
\def \b {\beta}
\def \a {\alpha}
\def \G {\Gamma}
\def \g {\gamma}
\def \c {\chi}
\def \h {\eta}
\def \s {\sigma}
\def \S {\Sigma}
\def \d {\delta}
\def \D {\Delta}
\def \e {\varepsilon}
\def \ep {\epsilon}
\def \r {\rho}
\def \o {\omega}
\def \O {\Omega}
\def \t {\vartheta}
\def \te {\theta}
\def \ta {\tau}
\def \m {\mu}
\def \n {\nu}
\def \z {\zeta}
\def \f {\varphi}
\def \F {\Phi}
\def \x {\xi}
\def \non {\nonumber}
\def \noi {\noindent}
\def \ra {\rightarrow}
\def \pr {\prime}
\def \im {\Rightarrow}
\def \eq {\Leftrightarrow}
\def \mau {\geqslant}
\def \miu {\leqslant}
\def \dps {\displaystyle}
\def \fr {\displaystyle\frac}
\def \sh {\slashed}
\def \wt {\widetilde}
\def\laq{~\raise 0.4ex\hbox{$<$}\kern -0.8em\lower 0.62
ex\hbox{$\sim$}~}
\def\gaq{~\raise 0.4ex\hbox{$>$}\kern -0.7em\lower 0.62
ex\hbox{$\sim$}~}
\def \lab {\label}
\def \fa {\forall}
\def \es {\exists}
\thispagestyle{empty} \vspace*{1cm} \rightline{BARI-TH/07-556}
\vspace*{2cm}
\begin{center}
  \begin{LARGE}
  \begin{bf}
$X(3872) \to D \overline D \gamma$ decays \\
 \vspace*{0.3cm}
and the structure of  $X(3872)$
 \vspace*{0.5cm}
  \end{bf}
  \end{LARGE}
\end{center}
\vspace*{8mm}
\begin{center}
\begin{large}
P. Colangelo$^a$,  F. De Fazio$^a$ and  S. Nicotri$^{a,b}$
\end{large}
\end{center}
\begin{center}
\begin{it}
$^a$Istituto Nazionale di Fisica Nucleare, Sezione di Bari,
Italy\\ $^b$ Universit\'a degli Studi di Bari, Italy
\end{it}
\end{center}
\begin{quotation}
\vspace*{1.5cm}
\begin{center}
  \begin{bf}
  Abstract\\
  \end{bf}
  \end{center}
\noindent
It has been suggested that the  radiative  $X \to D \overline D \gamma$ decay modes are useful
to shed light on  the structure of the  meson $X(3872)$, since the ratio
$ R={\Gamma(X \to D^+ D^- \gamma) \over \Gamma(X \to D^0 \overline D^0 \gamma)}$ is
expected to be small ($R \ll 1$) if $X$ is a molecular $D^{*0} \bar D^0$ state. We
 compute   $R$  in a   $\bar c c$ $J^{PC}=1^{++}$  description of $X$  finding that  it
is tiny in a wide range of hadronic parameters governing the decay.  A discrimination between the 
 molecular and  $\bar c c$ description  can be obtained through  the analysis of the photon spectrum.
 \end{quotation}
 
\newpage
\baselineskip=18pt \vspace{2cm}

The quark structure of the meson $X(3872)$ is a subject of 
 discussions  due to  the various puzzling aspects this
particle presents at a careful  scrutiny \cite{reviews}. The
resonance was discovered in the invariant mass distribution of $J/\psi
\pi^+ \pi^-$ mesons produced  in $B^\pm \to K^\pm X \to K^\pm J/\psi
\pi^+ \pi^-$ decays; it appeared as a narrow peak together with
the structure  corresponding  to the $\psi(2 ^3S_1)$ charmonium
level, with mass $M(X)=3872.0 \pm 0.6 \pm 0.5$ MeV and  width
smaller than the experimental resolution:  $\Gamma(X)<2.3$ MeV
($90\%$ C.L.) \cite{Choi:2003ue}. Confirmation of the state in $B$
decays was obtained later on
  \cite{Aubert:2004ns}, after the observation of  the structure in $p \bar p$ collisions
 at the Tevatron   with mass
 $M(X)=3871.4 \pm 0.7 \pm 0.4$ MeV \cite{cdf} and
 $M(X)-M(J/\psi)=774.9\pm 3.1 \pm 3.0 $ MeV  \cite{d0},  and width consistent with the detector
 resolution. The
$\pi^+ \pi^-$ spectrum  displayed a maximum  in the region of large invariant mass \cite{Choi:2003ue,Aubert:2004ns,pipispectrum}.

The meson $X(3872)$, whose average values of  resonance parameters
quoted  by the Particle Data Group  2006  are $M(X)=3871.2 \pm
0.5$ MeV  and  $\Gamma(X)<2.3$ MeV ($90\%$ C.L.) \cite{PDG}, was
not   observed in $e^+ e^-$ annihilation; moreover, searches
for charged partners, made by looking at the $J/\psi \pi^\pm \pi^0$
channel,   produced negative results
 \cite{chargedX}. The state was neither  found in  the $J/\psi \eta$ channel \cite{psieta}
 nor  in $\gamma \gamma$ fusion  \cite{gammagamma}.
 As for production in
 $B$ decays, the ratio
 $\displaystyle \frac{B(B^0 \to K^0 X)}{B(B^+ \to K^+ X)}=0.50 \pm 0.30 \pm 0.05$
 was measured \cite{babarkpi}.

On the basis of the observation of the radiative  mode $X \to J/\psi \gamma$ ,  with the measurement
$\displaystyle \frac{B(X \to J/\psi \gamma)}{B(X \to J/\psi \pi^+ \pi^- )}=0.19 \pm 0.07$
\cite{belle3p},
the  charge conjugation of the state is established: C=+1; moreover,
 the angular distribution of the final state is compatible with the spin-parity assignment
 $J^P=1^+$ (even though $2^-$ is not excluded)
  \cite{Abe:2005iy}, so that the most  likely  quantum number assignment for $X(3872)$
is $J^{PC}=1^{++}$.

 Together with these measurements, a  near-threshold  $D^0 \bar D^0 \pi^0$  enhancement in
 $B \to D^0 \bar D^0 \pi^0 K$ decay
was recently reported, with the peak at $M=3875.4 \pm 0.7 ^{+1.2}_{-2.0}$ MeV and
$B(B\to K X \to K D^0 \bar D^0  \pi^0)=(1.27 \pm 0.31^{+0.22}_{-0.39}) \times 10^{-4}$
\cite{Gokhroo:2006bt}.
If the enhancement is entirely due to $X(3872)$ one derives that
  $\displaystyle \frac{B(X \to D^0 \bar D^0  \pi^0)}{B(X \to J/\psi \pi^+ \pi^- )}=9\pm4$   \cite{BelleDoDopi},
therefore $X$  mainly  decays into final states with open charm mesons.
Notice that the central value of the mass measured in the
  $D^0 \bar D^0 \pi^0$ mode
  is $4$ MeV 
  higher than the PDG value (although with a large asymmetric systematic error  $\Delta M=+1.2,-2.0$ MeV).

These measurements, although not fully consistent with the expectations based on 
charmonium models (mainly as far as the mass of the state is concerned), do not contradict the
interpretation of $X(3872)$ as a $\bar c c $ state. However,
another hadronic  decay mode was observed for  $X(3872)$:
 $X \to J/\psi \pi^+ \pi^- \pi^0$  with  $\displaystyle \frac{B(X \to J/\psi \pi^+ \pi^- \pi^0)}{B(X \to J/\psi \pi^+ \pi^- )}=1.0 \pm 0.4 \pm 0.3$ \cite{belle3p,Y3930}.
 Presence of both  decay channels in two and three pions  implies G-parity violation or, if the
 two modes are
 considered as  induced by $\rho^0$ and $\omega$ intermediate states, isospin violation:
 this  suggested the conjecture
 that  $X(3872)$ is not a charmonium ($\bar c c$) state, but a hadron of more complex
  quark content. In the search of  the right interpretation, 
the   coincidence between the resonance mass  as 
 averaged by PDG   and the
 $D^{*0} \overline D^0$ mass: 
 $M(D^{*0} \overline D^0)=3871.2\pm 1.0$ MeV,   inspired  the proposal that
$X(3872)$  could be  a realization of the molecular quarkonium \cite{okun}, a bound state of two
 mesons $D^{*0}$ and $\overline D^0$
with small binding energy \cite{molec0,molec,interpretations},  an interpretation that would allow to account for a few
properties of $X(3872)$. For example,
 describing the wave function of $X(3872)$  through various hadronic components \cite{voloshin1}:
\be
|X(3872)>=a \, |D^{*0} \bar D^0+ \bar D^{*0}  D^0> + b \,  |D^{*+}  D^-+  D^{*-}  D^+> + \dots
\ee
(with $|b| \ll |a|$)
one could explain why this state seems not to have  definite isospin,  why
 the decay mode $X \to J/\psi \pi^0 \pi^0$ has not been found, and why, if the molecular binding
 mechanism is provided by a single pion exchange, there are no  $D \overline D$ molecular states:
 indeed no structures were  found in the  range of mass corresponding to $2 m_{D^0}$ or $2 m_{D^\pm}$. 
 Moreover, non observation of a  bound state of charged $D^{*+}D^-$  mesons can also be justified since a single pion exchange would produce a repulsive interaction in this channel 
 \cite{molec0}.

Noticeably, in the molecular interpretation the  resonance $X_b(10604)$ would be expected as a bound
state of $B$ and $B^*$; this resonance has not been observed, so far, so that the  prediction  deserves experimental investigations.  Moreover, it is also predicted that,
 since the decays of the $X(3872)$ resonance are mainly due to the decays of its meson components
 in case of peripheral transitions,
 the radiative decay in neutral $D$ mesons: $X \to D^0 \bar D^0 \gamma$ should be dominant with respect to
$X \to D^+  D^- \gamma$ \cite{voloshin1}.

The description of $X(3872)$ in a simple charmonium  scheme,  in which  it  would be identified  as the first radial excitation of the $J^{PC}=1^{++}$ state,
  presents alternative arguments  to the molecular description \cite{charmonium}.  
 A problem is that the molecular binding mechanism still needs to be clearly identified,  and the role of single $\pi^0$ exchange has to be further investigated.
\footnote{For example,  it was argued \cite{suzuki} that
  the molecular binding mechanism  cannot be  a single
$\pi^0$ exchange, since this would  produce an attractive  potential which  is a delta function
in space
and  therefore it would not give rise to a bound state. However, this argument is controversial: 
a detailed discussion can be found in the Appendix B  of the  first review in Ref.\cite{reviews}.} 
Concerning the isospin (G-parity) violation, in order to correctly interpret  the large value of the ratio
$\displaystyle \frac{B(X \to J/\psi \pi^+ \pi^- \pi^0)}{B(X \to J/\psi \pi^+ \pi^- )}$ one has to
consider that phase space effects in two and three pion modes are very different.
The ratio of the amplitudes is  smaller:
$\displaystyle \frac{A(X \to J/\psi \rho^0)}{A(X \to J/\psi \omega)}\simeq 0.2$, so that the
isospin violating amplitude is
20\% of the isospin conserving one, an
 effect that could be related to another isospin violating effect,  the mass difference between
neutral and charged $D$ mesons, considering  the contribution of  $DD^*$ intermediate states
to $X$ decays.  The  prediction
$\Gamma(B^0 \to X K^0) \simeq \Gamma(B^- \to X K^-)$,   based on  the charmonium description, is neither confirmed nor excluded   by
the available measurements.  Admittedly, the $\bar c c$ interpretation leaves unsolved  the issue of the
eventual overpopulation of the level corresponding to  the first radial excitations of $1^{++}$ $\bar c c$ states resulting from the possible assignment of these quantum numbers to another structure  observed by Belle Collaboration,
$Y(3930)$ \cite{Y3930};  however, this new resonance is still not  confirmed and its properties not
fully understood, so that
 the charmonium option for $X(3872)$ seems  not excluded, yet.  A warning comes from the
 $D^0 \bar D^0 \pi^0$ signal which, if  due to $X(3872)$, can contribute to settle the question of
 the coincidence of the $X$ and $D^0 \bar D^{*0}$ mass,   a relevant  issue since
a  $X(3872)$ above the $D^0 \bar D^{*0}$ threshold is  difficult to explain in a molecular picture.

In this note we  address a particular  aspect of
$X(3872)$, namely the suggestion   that  the observation of the dominance
of the process $X \to D^0 \bar D^0 \gamma$  with respect to $X \to
D^+  D^- \gamma$ could be interpreted as a signature
of the molecular structure of $X(3872)$     \cite{voloshin1}.  
Assuming that $X(3872)$ is an ordinary $J^{PC}=1^{++}$ charmonium
state,  together with a standard
mechanism for the radiative transition into charmed mesons, we obtain that the
ratio $\displaystyle R={\Gamma(X \to D^+ D^- \gamma)\over \Gamma(X
\to D^0 \overline D^0 \gamma)}$ is  small and in particular  it  is tiny in a wide
 range of the hadronic parameters governing the
decays, so that the ratio $R\ll 1$ seems  not peculiar of 
$X(3872)$ being  a molecular quarkonium.

\begin{figure}[b]
 \begin{center}  \includegraphics[width=0.4\textwidth] {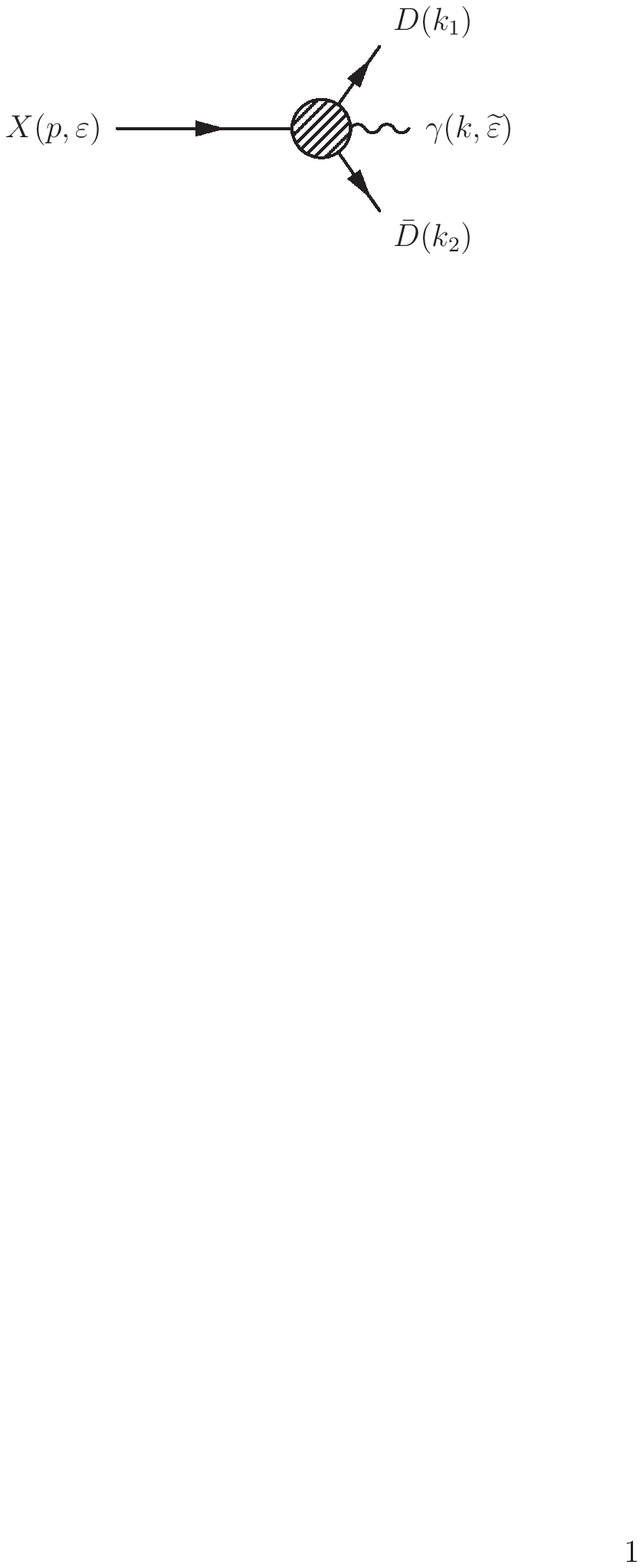}\\
\vspace*{-11.5cm}
\includegraphics[width=0.32\textwidth] {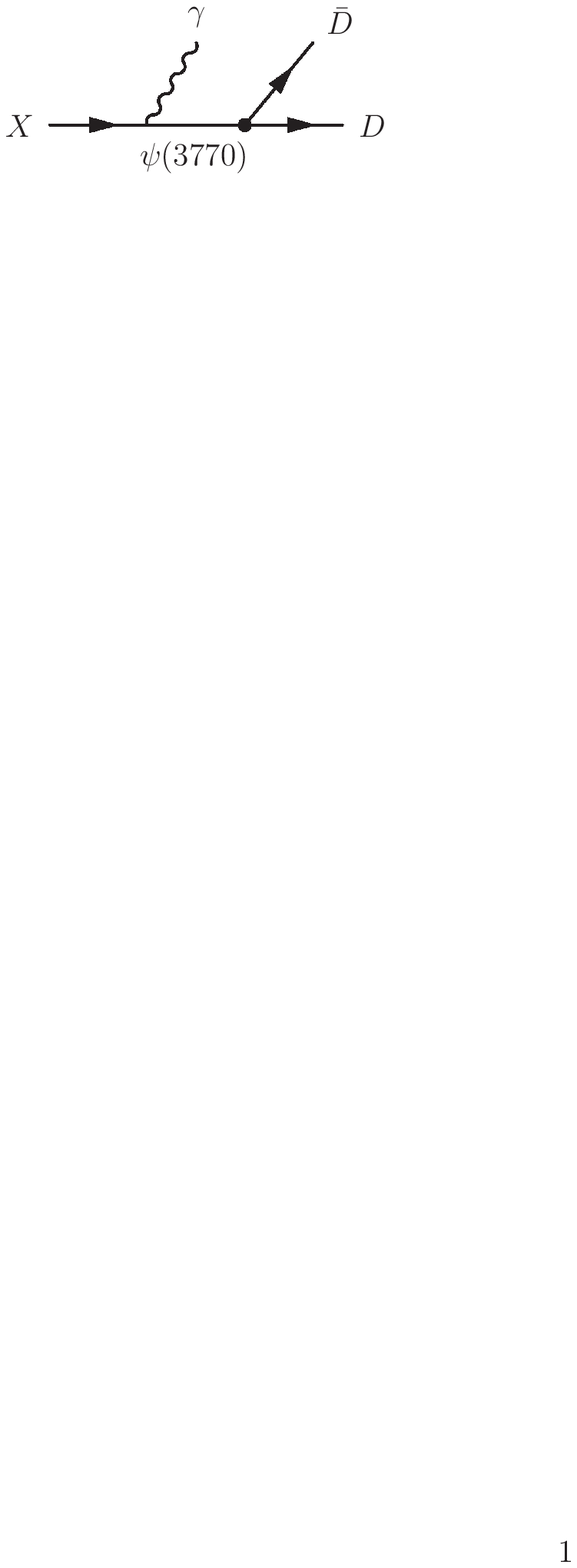}
\includegraphics[width=0.32\textwidth] {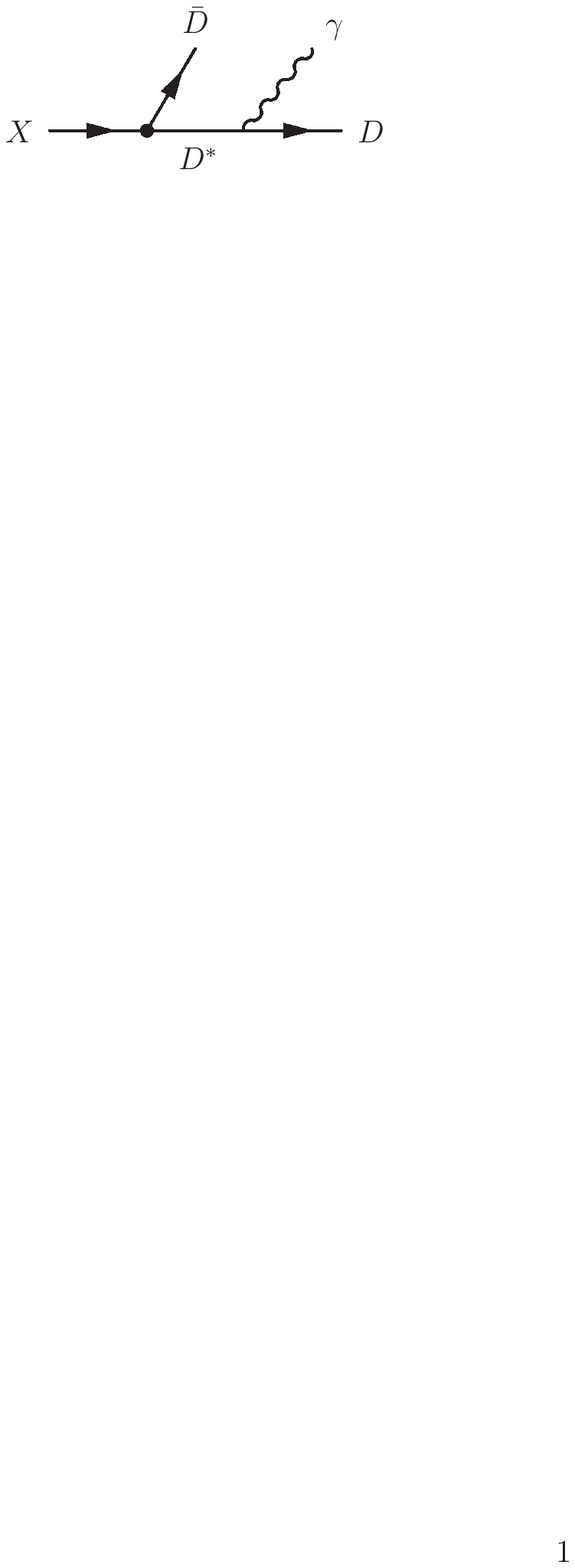}
\includegraphics[width=0.32\textwidth] {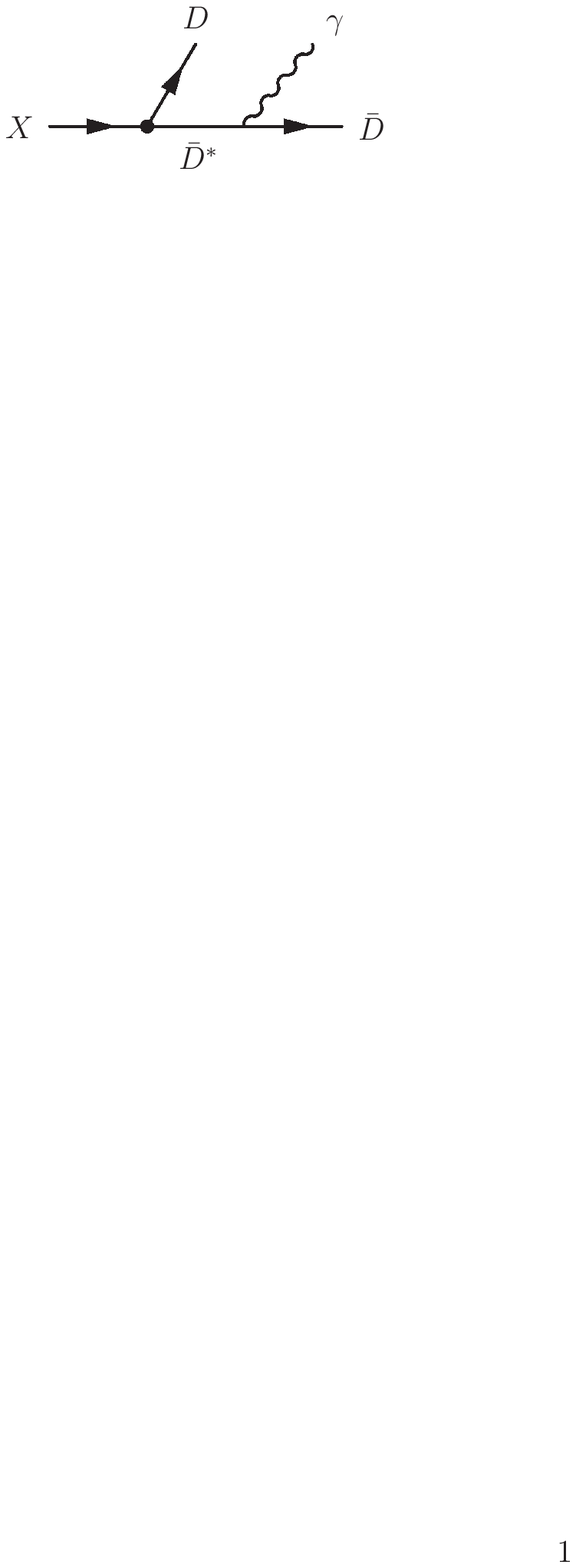}
 \vspace*{-11cm}
 \caption{Diagram describing the radiative  modes $X \to D \bar D \gamma$ (top),  and
 contributions  corresponding to the  intermediate states nearest to their mass shell (bottom).}
  \label{fig:gamma}
 \end{center}
\end{figure}
%
%
In order to study the transition $X(3872)(p,\epsilon)\to D(k_1)
\bar D(k_2) \gamma(k,\tilde \epsilon)$  ($p, k_1, k_2$ and $k$ are momenta, $\epsilon, \tilde \epsilon$
 polarization vectors)
we assume that the
radiative decay amplitude is dominated by pole diagrams with
intermediate particles nearest to their mass shell, as the  ones
depicted in fig.\ref{fig:gamma} which  involve $D^*$  and the
$\psi(3770)$  mesons as intermediate states. These
amplitudes can be expressed in terms of two  unknown  quantities:
 the coupling constant governing the $X \bar D D^*(D \bar D^*)$ matrix elements,  and the
 coupling appearing in the $X \psi(3770) \gamma$ matrix element, since 
 information about $D^*D\gamma$ and $\psi(3770)D\bar D$ couplings can be inferred from 
 experimental data.
 
For  the matrix element  $X \bar D D^*(D \bar D^*)$  we  use  a
formalism suitable to describe the interaction of the heavy
charmonium with  the doublet of heavy pseudoscalar and vector
meson states \cite{chiral}: the four states corresponding to the first
radial excitation of $\ell=1$ $\bar c c$ mesons,  which are
degenerate in the limit $m_c\to \infty$, can be described by the
multiplet: \be P^{(Q \bar Q)\mu}=\left( {1 + \spur{v} \over 2}
\right) \left( \chi_2^{\mu \alpha}\gamma_\alpha +{1 \over
\sqrt{2}}\epsilon^{\mu \alpha \beta \gamma} v_\alpha \gamma_\beta
\chi_{1 \gamma}+ {1 \over \sqrt{3}}(\gamma^\mu-v^\mu) \chi_0
+h_1^\mu \gamma_5 \right)\left( {1 - \spur{v} \over 2} \right)
\label{pwave} \ee where $\chi_2$, $\chi_1$ and $\chi_0$ correspond
to the spin triplet with $J^{PC}=2^{++}, 1^{++}$ and $0^{++}$, respectively, 
while the spin singlet $h_1$ has $J^{PC}=1^{+-}$.  In the $\bar c c$ interpretation
$X(3872)$ is described by $\chi_1$. The expression of the
multiplet is analogous to that describing the  lowest radial states, $\chi_{c0,1,2}$ and $h_c$; the fields
in eq.(\ref{pwave}) contain a factor $\sqrt m$, with $m$ the meson
mass. The strong interaction with the $D$ and $D^*$ mesons can be
described by the effective Lagrangian \cite{pham}
\be {\cal L}_1= i {g_1} Tr
\left[P^{(Q \bar Q)\mu} {\bar H}_{1a} \gamma_\mu {\bar H}_{2a}
\right] + h.c. \label{lagr1new} \ee 
where  the fields $H_{1,2}$
represent the spin doublets  ($D,D^*$) and ($\bar D, \bar D^*$),
respectively; $H_{1a}$ is the field describing the heavy-light
mesons with quark content $Q{\bar q_a }$  and four-velocity $v$,  $D^{(*)0}, D^{(*)+},
D_s^{(*)}$:
 \be H_{1a}=\left( {1 + \spur{v} \over 2} \right)
[M^\mu_a \gamma_\mu -M_a \gamma_5]  \,, \label{hq} \ee 
while
$H_{2a}$ describes the heavy-light mesons with quark content
$q_a{\bar Q}$, $\overline D^{(*)0}, D^{(*)-},  \overline
D_s^{(*)}$:
 \be H_{2a}= [M^{\prime \mu}_a \gamma_\mu - M^\prime_a
\gamma_5] \left( {1 - \spur{v}\over 2} \right) \label{h2} \ee 
with
$\bar H_{1,2}=\gamma^0 H_{1,2}^\dagger \gamma^0$. The effective
Lagrangian (\ref{lagr1new}) accounts for the fact that the two
heavy-light $D, D^*$ mesons are coupled to the charmonium state in
S-wave. Moreover, this expression is invariant under independent
rotations of the spin of the heavy quarks,  since these spins are
decoupled in the infinite heavy quark mass limit. Invariance under
heavy quark (antiquark) spin rotations can  be obtained
considering that under independent heavy quark spin
transformations:
 $S_1 \in SU(2)_{Q}$ and  $S_2 \in SU(2)_{\bar Q}$,
the following transformation properties hold for the various multiplets:
\bea
H_{1a} \to S_1 H_{1a} & \hskip 0.5 cm &
{\overline H}_{1a} \to  {\overline H}_{1a} S_1^\dagger \,\, \nonumber \\
H_{2a} \to H_{2a}S_2^\dagger & \hskip 0.5 cm &
{\overline H}_{2a} \to  S_2{\overline H}_{2a}\,\, \nonumber \\
P^{(Q \bar  Q)\mu} \to  S_1 P^{(Q \bar Q)\mu} &\hskip 0.5 cm &
P^{(Q \bar Q)\mu} \to P^{(Q \bar Q)\mu} S_2^\dagger \,\,\, .  \label{su2spin}
\eea

Using the effective Lagrangian (\ref{lagr1new})  
the couplings $XD^0{\bar D}^{*0}$ and $X{\bar D}^0
D^{*0}$ (or $X D^+{ D}^{*-}$ and $X{D}^- D^{*+}$)
 which enter in the calculation of the second and the
third diagrams in fig.\ref{fig:gamma},  respectively, can be expressed in terms of
the  constant $g_1$. For later
convenience, we  use  the dimensionless coupling constant $\hat g_1=g_1 \sqrt{m_D}$.
Due to  isospin symmetry, the couplings of the meson
$X$ to charged and neutral $D$ are equal, at odds with the molecular description where
$X$ mainly  couples to neutral $D$.

The second and third diagrams in fig.\ref{fig:gamma}  also require  the
knowledge of the electromagnetic vertex $ D^{*}  D \gamma$. We use
the  parametrization: 
\be 
<D(k_1)  \gamma(k,{\tilde \epsilon})|D^{*}(p_1,\xi)> =i \, e \, c^\prime \,
\epsilon^{\alpha \beta \tau \theta} \, {\tilde \epsilon}^*_\alpha
\, \xi_\beta \, p_{1 \tau} \, k_{\theta} \,, \label{d*dgamma}
\ee 
where the parameter $c^\prime$ accounts for the  contributions
of  the photon coupling to  both the charm and the light quark \cite{amundson}:
\be c^\prime={e_c \over m_c}+{e_q \over \Lambda_q} \,\,, \ee 
with $e_c$ and $e_q$  the charm and the light
quark charges in units of $e$, therefore 
$e_q=2/3 \, (-1/3)$ 
for neutral (charged)  charmed mesons. We use the value $m_c=1.35$ GeV for  the charm quark mass 
\cite{PDG};   $\Lambda_q$  can be fixed  from 
$D^*$ data  since,  using $\Gamma(D^{*+})=96 \pm 22 $
KeV and ${\cal B}(D^{*+} \to D^+ \gamma)=(1.6 \pm 0.4) \%$
\cite{PDG}, we  obtain  $\Lambda_q=335 \pm 29$ MeV. This also  implies, 
from   ${\cal B}(D^{*0} \to D^0 \gamma)=(38.1 \pm
2.9) \%$ \cite{PDG}, that  the   $D^{*0}$ width can be estimated as
$\Gamma(D^{*0})=102 \pm 16 $ KeV (the present
upper  bound  is $\Gamma(D^{*0})<2.1$ MeV \cite{PDG}).

Coming to the hadronic parameter $c$ governing the
radiative $X \psi(3770) \gamma$ matrix element and  entering in  the
first diagram in fig.\ref{fig:gamma}: \be
<\psi_{(3770)}(q , \eta) \gamma(k, \tilde \epsilon)|X(p,
\epsilon)>=i \, e \,c \, \epsilon^{\alpha \beta \mu \nu} \, {\tilde
\epsilon}^*_\alpha \, \epsilon_\beta \, \eta^*_\mu \, k_\nu \,\,\, ,
\label{xpsigamma} \ee 
this parameter  is also unknown.
On the other hand,   the coupling between
$\psi(3770) D {\bar D}$, which  appears in the expression  of
the first diagram in fig.\ref{fig:gamma}, is known from the experiment.
Using the definition: \be <D(k_1) {\bar
D}(k_2)|\psi(q,\eta)>=g_{\psi D\bar D} \, \eta \cdot k_1
\label{psidd-matrixel} \ee 
and the value
$\Gamma(\psi(3770))=23.0 \pm 2.7$ MeV \cite{PDG}, 
together with the observation that the  $\psi(3770)$ width is  saturated by
 $D \bar D$ modes, we obtain 
\be g_{\psi D \bar D}=25.7 \pm 1.5 \ee
 both for  charged and neutral $D$ meson pairs. Notice that in this determination we do not need to adopt  any interpretation for the $J^{PC}=1^{--}$ $\psi(3770)$ state, a meson the properties of which are still  
under scrutiny  \cite{voloshin3}. Another point to be stressed is that we determine the coupling constants
$g_{\psi D \bar D}$ and $c^\prime$  from on-shell processes and use them
in the vertices  in fig.\ref{fig:gamma} neglecting possible form-factor effects. Inclusion of form factors
would represent an additional source of theoretical uncertainty; however, in our case the intermediate
states are nearly on-shell, therefore form factor effects are expected to be small.

\begin{figure}[h]
\begin{center}
\includegraphics[scale=0.7]{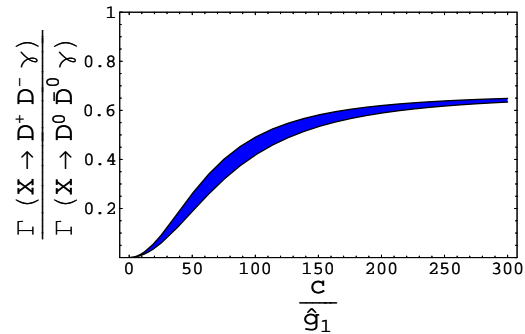}
\end{center}
\caption{\footnotesize{Ratio of charged $X \to D^+ D^- \gamma$ to
neutral $X \to D^0 \bar D^0 \gamma$ decay  widths  versus the ratio
of hadronic parameters $c/\hat g_1$.}}\label{fig:ratio} \vspace*{1cm}
\end{figure}

We can now evaluate the
ratio $\displaystyle R={\Gamma(X \to D^+ D^- \gamma) \over
\Gamma(X \to D^0 \overline D^0 \gamma)}$ as a function of the
ratio of the two  couplings $\displaystyle{c \over {\hat
g}_1}$ and  including  the   uncertainties on
$\Gamma(D^{*+})$,  $\Gamma(\psi(3770))$,
$\Lambda_q$ and  $g_{\psi D \bar D}$. The result is plotted in fig.\ref{fig:ratio}, 
where it is shown that  in any case 
$R < 0.7$. For large vales of
 $\displaystyle{c \over {\hat g}_1}$  the error
on $R$ is  small, since in this case  only $\psi(3770)$ contributes to the amplitudes.

\vspace*{1cm}

The result depicted in fig.\ref{fig:ratio} 
shows that there is always a suppression of the radiative $X$ decay mode into charged
$D$ mesons  with respect to the mode with
neutral $D$. Moreover, for small values of $\displaystyle {c \over \hat g_1}$  the ratio $R$ is tiny, so that
this is not peculiar of a molecular structure of $X(3872)$. The suppression of the contribution of the two
last diagrams in fig.\ref{fig:gamma} in case of charged $D$ is mainly due to the higher mass of 
$D^{*\pm}$ with respect to $D^{*0}$, an important effect in the kinematic conditions of the process.

\begin{figure}[t]
\begin{center}
\includegraphics[width=0.35\textwidth] {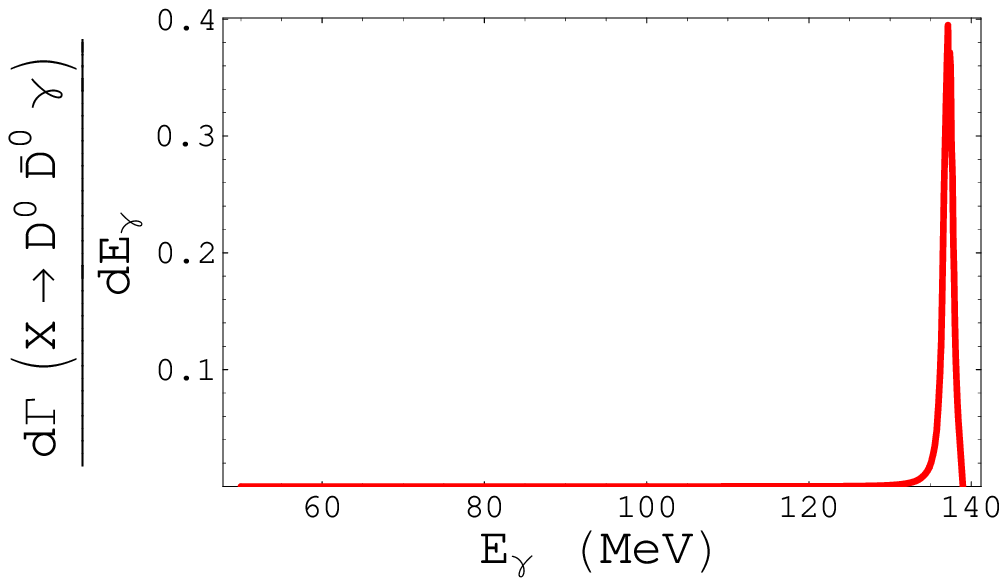} \hspace{0.35cm}
 \includegraphics[width=0.35\textwidth] {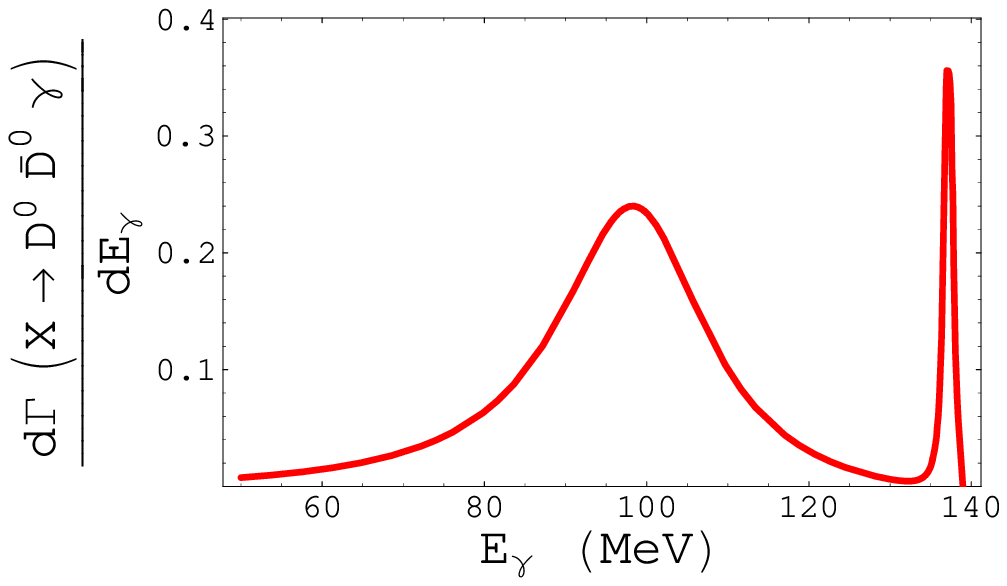}\\ \vspace{0.35cm}
 \includegraphics[width=0.35\textwidth] {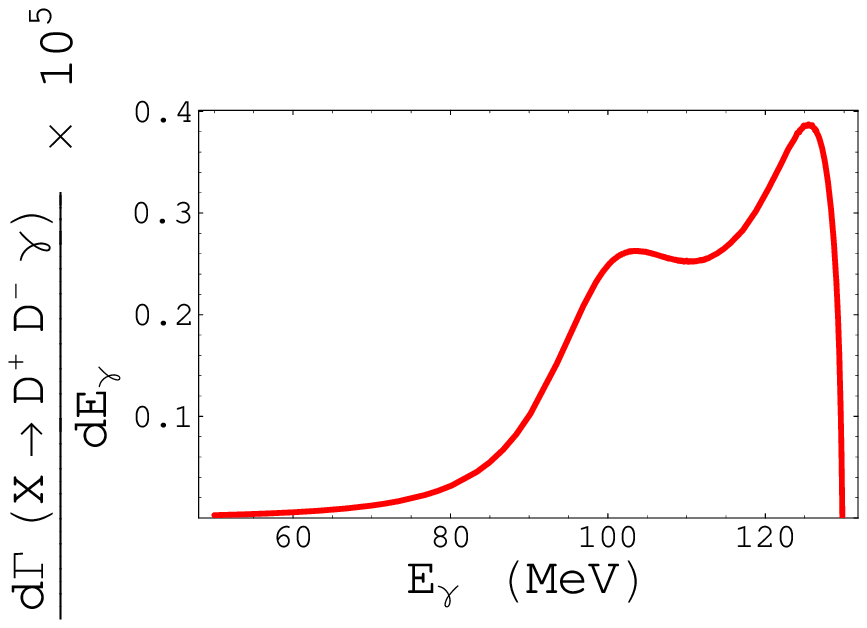} \hspace{0.35cm}
 \includegraphics[width=0.35\textwidth] {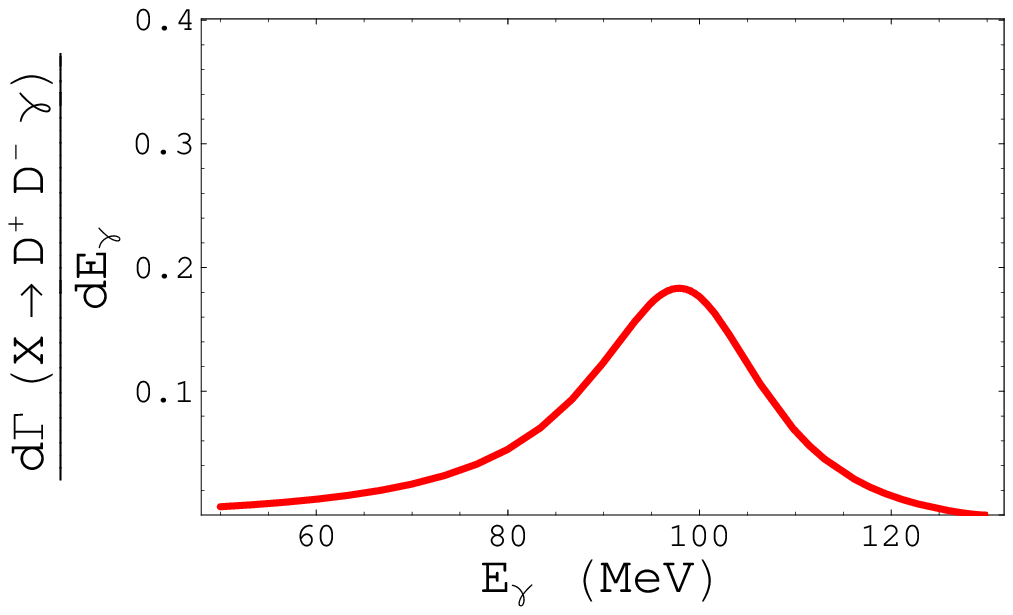}
\end{center}
\caption{\footnotesize{Photon spectrum (in arbitrary units) in
$X \to D^0 \bar D^0 \gamma$ (top) and  $X \to D^+ D^- \gamma$ (bottom)
decays  for values of the hadronic parameter $c/\hat g_1=1$ (left) and $c/\hat g_1=300$ (right).}}\label{fig:spectra} \vspace*{1cm}
\end{figure}

The photon spectrum in radiative $X$ decays to both neutral and charged
$D$ meson pairs for two representative values of  $\displaystyle{c \over \hat g_1}$,
namely $\displaystyle{c \over \hat g_1}=1$ and $\displaystyle{c \over \hat g_1}=300$, is depicted in fig.\ref{fig:spectra}.
For low value of the parameter $\displaystyle{c \over \hat g_1}$, i.e. in the condition where the intermediate $D^*$ dominates the decay amplitude, the photon spectrum in the $D^0 \bar D^0 \gamma$ mode  
essentially coincides with the line corresponding to the $D^*$ decay at $E_\gamma \simeq 139$ MeV and width determined by the $D^*$ width. The narrow peak is different from the line shape expected
in a molecular description, which is related to the wave function of the two heavy mesons bound in the $X(3872)$,  in particular  to the binding energy of the system, being broader for larger binding energy. On the other hand, the photon spectrum in the charged $D^+  D^- \gamma$ mode is broader,
with a peak at  $E_\gamma \simeq 125$ MeV, the total  $X \to D^+  D^- \gamma$ rate being severely suppressed with respect to the   $X \to D^0  \bar D^0 \gamma$ one.

At the opposite side of the $\displaystyle{ c \over \hat g_1}$  range,  where $\psi(3770)$ gives a large contribution to the radiative amplitude, a peak at $E_\gamma \simeq 100$ MeV appears  both in neutral and charged
$D$ meson modes, in the first case together with  the structure at  $E_\gamma \simeq 139$ MeV. This spectrum was described also in \cite{voloshin1}, where in this case the radiative decay was interpreted as deriving from the $\bar c c$ core of  $X(3872)$. In this range of parameters  the ratio of the 
  $X \to D^+  D^- \gamma$ to  $X \to D^0  \bar D^0 \gamma$ rates reaches the largest value. 

The experimental determination of the photon spectrum of the type depicted in fig.\ref{fig:spectra}, together with the measurement of the $X \to D \bar D \gamma$ widths
 is a challenging task. Nevertheless, this  measurement  is   important 
to  shed light on the structure of $X(3872)$.

\begin{figure}[h]
\begin{center}\includegraphics[width=0.32\textwidth] {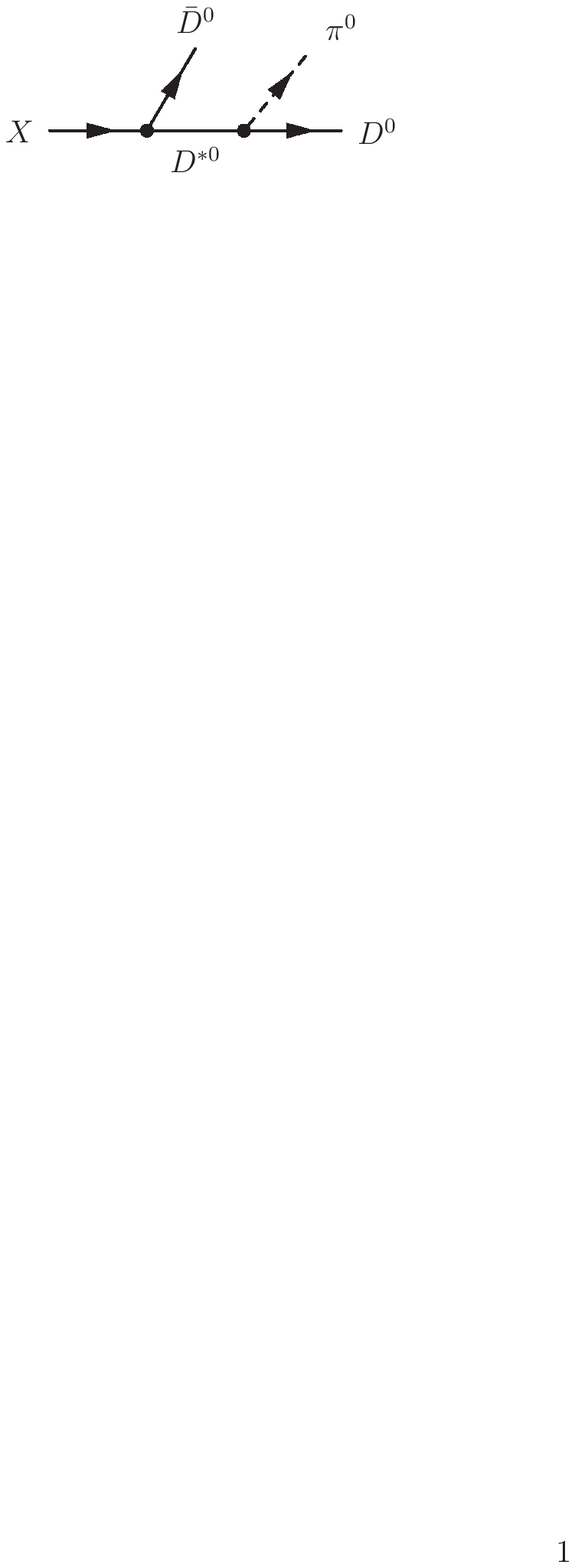}
\includegraphics[width=0.32\textwidth] {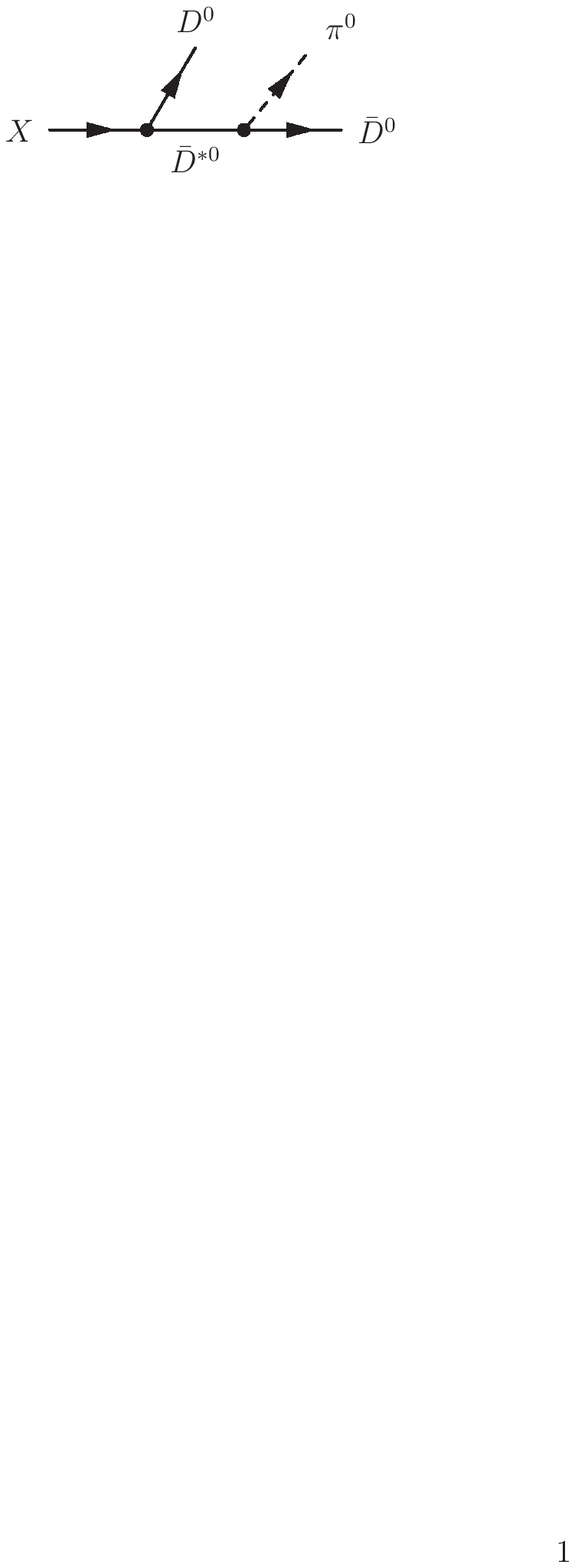}
\end{center}
\vspace{-13cm} \caption{\footnotesize{Diagrams contributing to the
mode $X \to D^0 {\bar D}^0 \pi^0$}.} \label{fig:pion} \vspace*{1cm}
\end{figure}

\begin{figure}[t]
\begin{center}
\includegraphics[scale=0.7]{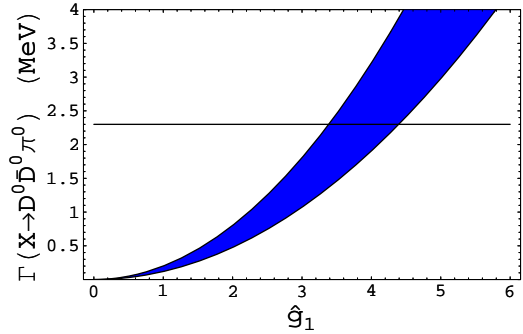}
\end{center}
\caption{\footnotesize{The width $\Gamma(X(3872) \to D^0 {\bar D}^0 \pi^0)$ versus the coupling constant
${\hat g}_1$  obtained  assuming that the decay proceeds as  in fig.\ref{fig:pion}
and  using the central values of $M(X(3872))$ and $M(D^0)$.
The horizontal line corresponds to the present bound on $\Gamma(X(3872))$.
}}\label{fig:gammapi} \vspace*{1cm}
\end{figure}

Information on the hadronic parameter
${\hat g}_1$ can be gained through  the mode $X(3872)\to
D^0\bar D^0 \pi^0$   described by
 pole diagrams  such as those in fig.\ref{fig:pion}.
%
%
The  needed new quantity  with respect to the radiative decay  is the
coupling constant $D^*D\pi$, which can be extracted from experimental data.
We define: \be <D^0 (k_1)
\pi^0(k)| D^{*0}(p_1, \xi) >={\sqrt{2 m_{D^0} m_{D^{*0}}} \over f_\pi} \, g \label{gd*dpi}
\ee 
with $f_\pi$ the pion leptonic constant and  the coupling $g$ 
 identified with the universal constant
governing the interaction of $J^P=(0^-,1^-)$ heavy-light mesons with
light pseudoscalar mesons in the heavy quark and chiral limit
\cite{chiral}. Using the present determination of $\Gamma(D^{*+})$ together with  the  branching fractions 
 ${\cal B}(D^{*+} \to D^0 \pi^+)=(67.7 \pm 0.5) \, \%$ and  
  ${\cal B}(D^{*+} \to D^+ \pi^0)=(30.7 \pm 0.5) \, \%$   \cite{PDG}
we obtain
$g=0.64 \pm 0.07$ and $g=0.60 \pm 0.07$, respectively.
\footnote{This value for the $D^*D\pi$ coupling is larger than obtained by various
methods, for example in ref.\cite{gddpi}; it comes from    the $D^{*+}$ width
currently quoted by PDG \cite{PDG} and determined by a single measurement in
 \cite{cleog}.}
This information would allow us to constrain ${\hat g}_1$ from the upper bound on
$\Gamma(X \to D^0 {\bar D}^0 \pi^0)$, since $\Gamma(X\to D^0 {\bar D}^0 \pi^0) < \Gamma(X(3872))<2.3 \,$ MeV.
Using  the central values of  the masses of $X(3872)$ and  $D^0$ we obtain ${\hat g}_1<4.5$, as
shown  in fig.\ref{fig:gammapi}:   therefore, a value of ${\hat g}_1$
of the typical size of the hadronic couplings can reproduce the small width of $X(3872)$, thus explaining
 one of the puzzling aspects of  the meson which are difficult to understand, for instance, in 
 a multiquark picture.
However, the numerical result for ${\hat g}_1$ critically depends on the meson masses, since
the phase space available for the process $X\to D^0 {\bar D}^0 \pi^0$ is tiny and the mass effects are essential.  Reducing the available phase space 
by considering the present  uncertainties on $M(X(3872))$ and $M(D^0)$  the  upper bound
for ${\hat g}_1$  is larger by about an order of magnitude, but still it has a size that could be expected for a typical hadronic coupling. 

To conclude,  our study  is based on a particular interpretation of $X(3872)$ and not on a determination of various 
hadronic parameters that can be done, e.g., in versions of the quark model.  Since
at present 
the  charmonium option for $X(3872)$ cannot be simply excluded,   the analysis of the photon
spectrum of  radiative  $X \to D \bar D \gamma$ decays can be useful in clarifying the situation.  
The  confirmation of the existence and of the properties of the resonance $Y(3930)$ reported by Belle Collaboration,  and a measurement with high precision
of the $X(3872)$ mass  from the $D^0 \bar D^0 \pi^0$ decay mode would  provide us with new important information, while, from the theory view point,  further studies of 
mechanisms for molecular binding  are  required.
Due to the importance of   demonstrating the existence of a
 hadronic configuration comprising  two bounded heavy mesons, such new investigations are worth
carrying out.

\vspace*{1cm}
\noindent{\bf Acknowledgments}
\noindent This work was supported in
part by the EU Contract  No. MRTN-CT-2006-035482, "FLAVIAnet".


\begin{thebibliography}{99}
\bibitem{interpretations}
Other proposals based on a multiquark picture are described in Ref. ~\cite{reviews}.
\end{thebibliography}
\end{document}